\documentclass[preprint,prd,tightenlines,nofootinbib,superscriptaddress]{revtex4-1}

\usepackage{graphicx,amsmath,amssymb,epsf, fdag, bm}

\usepackage{color}
\usepackage{graphicx}
\definecolor{red}{rgb}{1,0,0}
\def\lesssim{\ \hbox{\raise 2pt \hbox{$<$} \kern -13pt
                     \lower 3pt \hbox{$\sim$}}\ }
\def\greatersim{\ \hbox{\raise 2pt \hbox{$>$} \kern -13pt
                     \lower 3pt \hbox{$\sim$}}\ }
\input epsf.tex
\def\desepsf(#1 width #2){\epsfxsize=#2 \epsfbox{#1}}

\begin{document}

\hspace*{11.5 cm} {\small IFT-UAM/CSIC-12-39}

\hspace*{11.5 cm} {\small LPN12-050}

\hspace*{11.5 cm} {\small OUTP-12-08-P}

\vspace*{1.4 cm} 

\title{Forward $Z$-boson  production  and the   unintegrated  sea quark  density}
\author{F.\ Hautmann} 
\affiliation{Theoretical Physics, 
University of Oxford,    Oxford OX1 3NP}
\author{M.\ Hentschinski}
\affiliation{Instituto de F{\' i}sica Te{\' o}rica  
UAM/CSIC,    Universidad Aut{\' o}noma de Madrid, E-28049 Madrid}
\author{H.\ Jung}
\affiliation{Deutsches Elektronen Synchrotron, D-22603 Hamburg}
\affiliation{CERN, Physics Department, CH-1211 Geneva 23}

\begin{abstract}
Drell-Yan production  in the forward region at the Large Hadron 
Collider is  sensitive to multiple  radiation of  QCD  partons 
not collinearly ordered, emitted  over large rapidity intervals. 
We propose a method  to take  account of these radiative 
contributions   via a factorization 
formula which depends on the  unintegrated, or transverse momentum dependent, 
splitting function associated with  the  evolution  of the initial-state sea quark 
distribution. 
We analyze this formula numerically,  and point out kinematic effects 
from the initial-state transverse momentum on the vector boson spectrum.  
\end{abstract}


\maketitle

\section{Introduction} 

Many  aspects  of     the experimental program at the 
Large Hadron Collider (LHC)  
depend on  the analysis of  processes at large 
momentum transfers containing multiple hard scales. As such, they 
will   be influenced  by improved 
  formulations  of factorization  in QCD  
     at unintegrated level~\cite{jcc-book}, which serve  in multi-scale processes  
     to   control  perturbative  large 
logarithms to higher orders of perturbation theory  and 
  to  describe appropriately nonperturbative physics in the initial 
   and final states of the collision.   These   formulations  involve 
 transverse-momentum dependent  (TMD),  or   unintegrated,     
parton decay  and parton  density   functions (pdfs)~\cite{mert-rog}.       

A broad class  of  such multiple-scale  events 
   is given by  small-$x$ processes.  
 These  processes are one  of   the main sources  of  final states  
in the   central region at the LHC~\cite{ajaltouni}.   Besides,   
they  are responsible for the  
sizeable   rates  of       forward  large-p$_\perp$  jet  production at the LHC, 
giving rise to new    phenomenology in the LHC forward region 
compared to previous 
 collider  experiments,   e.g. forward jet physics~\cite{fwdjetphys1,fwdjetphys2}. 
Small-$x$  TMD factorization  and pdfs    
   (see for instance recent reviews in~\cite{avsar11,hj_rec})  
serve for the perturbative   resummation of  
high-energy logarithmic corrections 
and for the  development  of  parton shower  algorithms 
capable of  incorporating   multi-gluon  coherence  including 
the effects of large-angle, noncollinear  emission.

Most of  TMD computational tools 
relevant to small-$x$ and forward physics at the LHC  
have so far been developed   within 
a quenched approximation in which only gluon and valence quark 
contributions  are taken    into   account  
 at TMD level~\cite{fwdjetphys2,avsar11,hj_ang}.  
While  this gives a sensible approximation, 
based on the dominance of  spin-1  exchange processes  at high energies, 
 to  the   
asymptotic behavior  of  production 
processes  coupled to gluons such as heavy flavor and scalar production, 
it is mandatory to go beyond this approximation 
to  include  preasymptotic  effects,   
and to  treat final states associated with 
quark-initiated processes such as Drell-Yan production. 
  In this work we take first steps  to 
address  this  issue  by including sea quark  contributions and 
examine  forward Drell-Yan production.

Drell-Yan processes at the LHC are  instrumental 
 in  precision electroweak  measurements, 
in luminosity monitoring  and pdf determinations,   
 and in new physics searches~\cite{ajaltouni}. 
    Studies of Drell-Yan at TMD level have  
   recently 
  been performed  in the 
framework of the soft-collinear effective theory~\cite{becher-neu,man-petrie,idi-sci}.  
Results     on high-energy  logarithmic  
corrections to   Drell-Yan  have  been  obtained in 
    the  $ q g^* $    channel~\cite{Marzani:2008uh,Lipatov:2011sd}     
    and in the associated production  channel 
     $Z / W  + $ heavy quarks~\cite{michal08}.  
On the other hand,  early  attempts~\cite{kwie03,kimber,Watt:2003vf,hoeche08}  to 
 treat the unintegrated pdf evolution 
  beyond the quenched approximation  include  quarks 
    via   splitting probabilities to lowest order of perturbation 
  theory,     neglecting  
   any transverse momentum  dependence in the branching.  
In~\cite{Martin:2009ii}  k$_\perp$-dependent 
kinematic corrections are   included, while the  
 splitting  kernels  are still taken in lowest order. 
Also,  a program to  perform shower Monte Carlo evolution 
at unintegrated level has recently been proposed~\cite{jadach09}   
based on the expansion~\cite{CFP} in 
two-particle irreducible (2PI)  kernels.  This program is formulated  
 at the next-to-leading  logarithmic order. However,  this     
does not include  small-$x$  logarithmic effects    which  are present 
beyond  NLO  in  flavor singlet   distributions.

The approach of the present paper 
is based on  the high-energy form of the 
(off-shell)  TMD  quark Green function  introduced in~\cite{Catani:1994sq}.  This    
is obtained by    generalizing  to finite transverse momenta, in the high-energy 
region,   the 2PI expansion~\cite{CFP}.   
The main point     is to 
construct   unintegrated  sea-quark distributions    
 incorporating  the effects of   the TMD   gluon-to-quark 
 splitting kernel~\cite{Catani:1994sq},   which 
  contains   all single-logarithmic 
small-$x$  corrections to  sea quark evolution 
 for    any  order  of  perturbation theory.    
In order to  relate  this  parton  splitting  kernel to 
 forward vector boson production, we  analyze   the 
  flavor exchange process at high energy according to the 
``reggeized quark"  calculus~\cite{Bogdan:2006af, Lipatov:2000se}.  
This   extends 
 the effective action formalism~\cite{Lipatov:1995pn},  
 currently explored at NLO~\cite{Hentschinski:2011tz}, to  
amplitudes with quark  exchange 
in   terms of effective degrees of freedom, the so-called 
 reggeized quarks~\cite{Fadin:1976nw,knie-sale}. 
 The use  of the effective vertices~\cite{Bogdan:2006af, Lipatov:2000se}   
 ensures  gauge  
invariance of the  coefficients relevant to perform 
  the high-energy 
   factorization~\cite{Catani:1994sq,ktfac}  for vector boson production, 
 despite  the off-shell  parton. 
  
We then examine kinematic effects related to the difference between 
the virtuality of the exchanged parton  and  its transverse part 
 which, although formally subdominant  
both in the collinear and in the high-energy expansions,  can 
nevertheless be numerically non-negligible.  
These kinematic effects are similar to those  observed 
 in the study of~\cite{Martin:2009ii}.  
We perform  a numerical study 
of the kinematic terms  at  the level of the 
partonic $Z$ cross section, and  compare this  
 with the $ q g^*  \to  Z q $  result~\cite{Marzani:2008uh}.  

The  formulation  proposed   in this work 
  can be implemented  in a parton shower    Monte Carlo event generator.  
      Details  of   the formulation and    shower  implementation 
    will be described elsewhere~\cite{prepar}. 
The outline of  the present paper  is the following. In Sec.~2  
we  apply  the reggeized quark calculus  to   examine    $Z$ boson production 
and reobtain  the TMD  quark  splitting function. In Sec.~3 we   investigate 
kinematic effects in the factorization formula  and study this 
numerically.   We give conclusions in Sec.~4. 

\section{Reggeized quark and unintegrated quark  density}

In this section  we relate the $Z$ boson production cross section to the 
unintegrated quark density defined at high energy 
via  the  transverse momentum dependent  kernel~\cite{Catani:1994sq}. 
For definiteness we  consider factorization
of the $g^*q \to Zq$ matrix element into off-shell point-like $qq^* \to Z$ coefficient and
off-shell  gluon-to-quark splitting function, which is of direct phenomenological 
relevance for forward Drell-Yan production.

\begin{figure}[th]
  \centering
  \parbox{3.1cm}{\center \includegraphics[width=3.1cm]{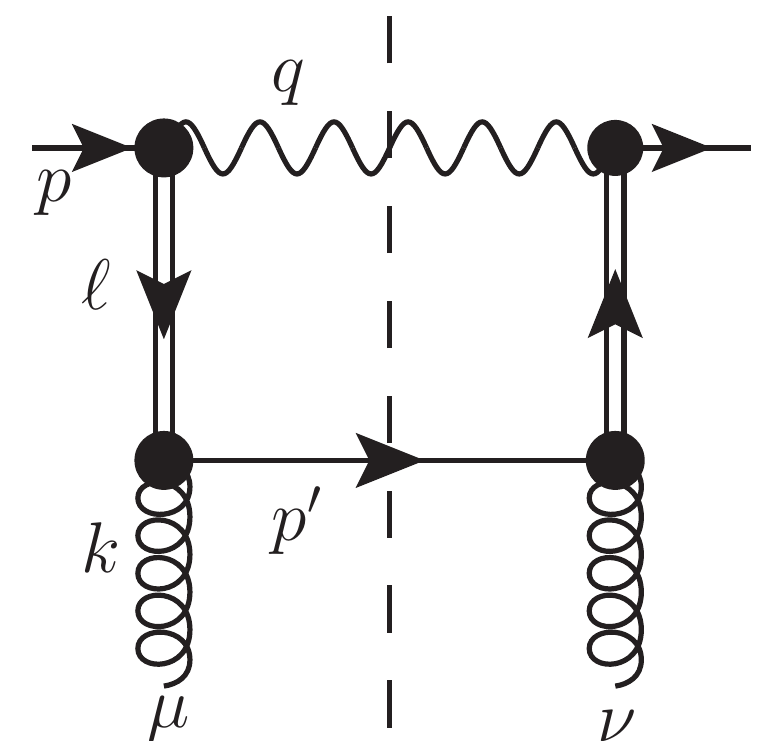}}
  \caption{\small The $g^*q \to Zq$ process within the reggeized quark
    formalism. Double lines with arrow indicate the effective
    reggeized quark exchange in the $t$-channel.}
  \label{fig:DY_reggeized}
\end{figure}

Within the reggeized quark formalism of~\cite{Bogdan:2006af,
  Lipatov:2000se} this process can be described in the high energy
limit by the single effective diagram  in fig.~\ref{fig:DY_reggeized}.   
The
double lines (with fermionic arrows) in the diagram represent the
exchange of off-shell (reggeized) quarks, while the black dots represent their effective 
couplings to $s$-channel quarks and gauge bosons~\cite{Bogdan:2006af}.  
We use the following Sudakov decomposition  for the initial and final momenta, 
\begin{align}
  \label{eq:Sudakov}
  p & = x_2 p_2 & q & = z x_1 p_1 + \left( x_2 + \frac{{t} + {\bm q}^2}{z x_1 s  }\right) p_2 + {\bm q} \notag 
\\
 k & = x_1 p_1 + {\bm k} & p'& = (1-z) x_1 p_1 - \frac{{t} + {\bm q}^2}{z x_1 s  } p_2  + {\bm k} - {\bm q} ,
\end{align}
where  $p_1$ and  $p_2$ are 
light-like four-momenta       respectively    
in the plus and minus lightcone directions,  $2
p_1\cdot p_2 = s$, and $\ell^2 = {t} $. 
The reggeized quark  propagators are purely
transverse fermion propagators, supplemented with a projector in
longitudinal momentum space,
\begin{align}
  \label{eq:RQprop}
  \parbox{1cm}{ \includegraphics[height=1.7cm]{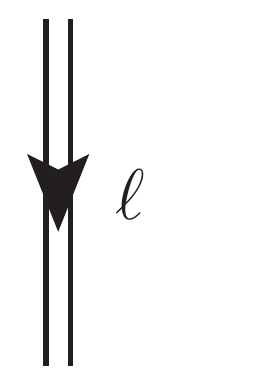}} &=  \frac{\fdag{p}_2 \fdag{p}_1}{2 p_1 \cdot p_2} \cdot \frac{i \cdot \fdag{\bm \ell}}{- {\bm \ell}^2+ i\epsilon }
&
\parbox{1cm}{ \includegraphics[height=1.7cm]{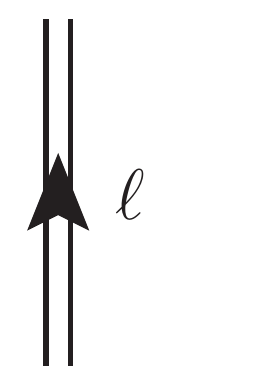}} &=  \frac{\fdag{p}_1 \fdag{p}_2}{2 p_1 \cdot p_2} \cdot \frac{i \cdot \fdag{\bm \ell}}{- {\bm \ell}^2+ i\epsilon }
\end{align}
The couplings of the off-shell quarks to $Z$ bosons and usual quarks
are
\begin{align}
  \label{eq:Zvertex}
  \parbox{3cm}{\center \includegraphics[height=1.7cm]{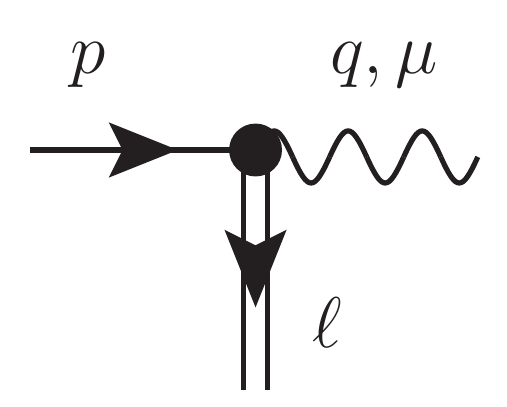}} &= 
\frac{i e}{\cos \theta_W \sin \theta_W} \Gamma^\mu_+(p, \ell, q) \left(V_f - A_f \gamma_5 \right)
,
\end{align}
with the ``plus" off-shell reggeized vertex given by  
\begin{align}
  \label{eq:GammaZ}
  \Gamma^\mu_+(p, \ell, q) =  \left( \gamma^\mu + \fdag{\ell} \cdot \frac{p_1^\mu \,}{p_1 \cdot q}   \right) \bigg|_{\ell \cdot p_1 = 0}   , 
\end{align}	
where the light-cone $p_2$ component of the $t$-channel momentum $\ell$ is set to zero. 
Similarly,  the lower vertex is given by 
\begin{align}
  \label{eq:gluonvertex}
  \parbox{2cm}{\center \includegraphics[height=2.1cm]{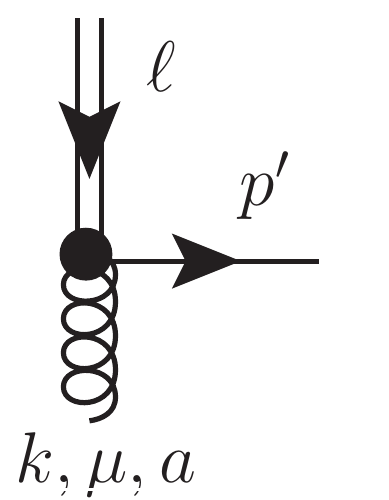}} &=
ig t^a \Gamma^\mu_-(k, \ell, p')    
\end{align}
with
\begin{align}
  \label{eq:Gamma_minus}
  \Gamma^\mu_-(k, \ell, p') &= \left( \gamma^\mu 
  +  \fdag{\ell} \cdot \frac{p_2^\mu \,}{p_2 \cdot k}   \right) \bigg|_{\ell \cdot p_2 = 0}   , 
\end{align}
where the  light-cone  $p_1$ component of the $t$-channel momentum is set to
zero.

The kinematic conditions $ \ell \cdot p_1 = 0$ and $\ell \cdot p_2 = 0$ 
 in eqs.~\eqref{eq:GammaZ} and~\eqref{eq:Gamma_minus}  
reflect  the strong ordering in the 
minus and plus lightcone  momenta~\cite{Bogdan:2006af,Lipatov:1995pn} and are 
 natural from the 
 point of view of high-energy factorization.  
 Using these conditions,  the 
 expression which results from
fig.~\ref{fig:DY_reggeized} agrees precisely with the high-energy   
expansion of the  $g^*q \to Zq$ matrix element, as  carried out 
in~\cite{Marzani:2008uh}   along the lines  of the high-energy 
 resummation~\cite{Catani:1994sq}.  
 However, the   expression   resulting from 
 the strong ordering  
 kinematics  in eqs.~\eqref{eq:GammaZ} and~\eqref{eq:Gamma_minus}  
 only   yields    
 a good approximation in the limit of asymptotically  large
partonic center-of-mass energy.   In the present analysis, we concentrate on  a 
 different aspect.  On one hand, we are interested in 
 improving  the  description for center-of-mass energies 
  which are not asymptotically  large. On the other hand,      we are interested in  
  obtaining   an off-shell (i.e., $k_T$-dependent)   factorization   
 in terms of the transverse-momentum dependent sea-quark distribution 
  defined  in~\cite{Catani:1994sq} from the two-particle irreducible expansion.  

To this end, 
we will relax some   of these  kinematic approximations. 
In the rest of this section   we  match
the high-energy factorized expression to  collinear
factorization by relaxing the condition $\ell \cdot p_2 = 0$.  In Sec.~3 
  we    discuss the effect of  relaxing the condition $\ell 
\cdot p_1 =0$.    
 Instead of   
eq.~\eqref{eq:Gamma_minus} we consider the effective vertex with exact
kinematics
\begin{align}
  \label{eq:gamma_tilde}
 \tilde{ \Gamma}^\mu_- (k,\ell, p' ) &= 
ig t^a \left( \gamma^\mu + \frac{p_2^\mu \,}{p_2 \cdot k}\fdag{\ell} \right) & \text{with}&
& \ell& = - z x_1 p_1 -   \frac{\hat{t} + {\bm q}^2}{z x_1 s  }    p_2  -  {\bm q}.
\end{align}  
Current conservation is satisfied~\cite{Bogdan:2006af,Lipatov:1995pn}    
 also  in  this more general case,    
\begin{align}
  \label{eq:minusvertex_gen}
  k_\mu  \tilde{\Gamma}^\mu_- (k,\ell, p') u(p') =  ig t^a \fdag{p}' u(p')  = 0.
\end{align}
In order to combine the above vertices with 
gluonic $k_T$ factorization~\cite{ktfac},    
we contract the above vertex with the transverse gluon 
momentum, 
\begin{align}
  \label{eq:gamma_tilde_regg}
 \tilde{ \Gamma}^\mu_- (k,\ell, p' ) \cdot \frac{{\bm k}_\mu}{\sqrt{{\bm k}^2}} .
\end{align}
This yields  the $q^*qg^*$-vertex which will be used in the following. 
  Owing to  eq.~\eqref{eq:minusvertex_gen},   
 this    is merely a rewriting from 
the longitudinal polarization 
  naturally associated with reggeized gluons  to  the transverse  
  polarization, which allows one  to perform the matching with the 
  collinear factorization~\cite{CFP,Catani:1994sq}.   
 Removing the condition $\ell \cdot p_2 =0$ for the  effective vertex,  we     also 
   promote the  
propagators  in eq.~\eqref{eq:RQprop}   to full four-dimensional propagators 
 as  follows,  
\begin{align}
  \label{eq:new_tchannelprop}
\parbox{1cm}{ \includegraphics[height=1.7cm]{RQ_prop1ell.pdf}}   &= 
  \frac{i \cdot \fdag{ \ell}}{ { \ell}^2+ i\epsilon }  \cdot  \frac{\fdag{p}_1 \fdag{p}_2}{2 p_1 \cdot p_2} &
\parbox{1cm}{ \includegraphics[height=1.7cm]{RQ_prop2ell.pdf}} &=  \frac{\fdag{p}_1 \fdag{p}_2}{2 p_2 \cdot p_1} \cdot \frac{i \cdot \fdag{ \ell}}{ { \ell}^2+ i\epsilon }.
\end{align}
As in~\cite{CFP,Catani:1994sq}, 
to perform  the  matching in the collinear region 
we  include   an upper bound 
on  four-momentum exchanged in 
the $t$-channel, $ \ell^2  <  \mu_F^2$,  with 
$\mu_F$ the factorization scale.

Using  the setup  described above,  we find that 
the $k_T$ factorized $qg^* \to
Zq$ cross-section  can be  written  as   
\begin{align}
  \label{eq:heXsec}
   \sigma^{k_T-\text{fact.}}_{qg^* \to Zq} &=   \int_0^1 d z \int \frac{d^2 {\bm q} }{ \pi }    \Theta\left(\mu_F^2 - \ell^2 \right)  \,   \hat{\sigma}(x_1x_2 s, M_Z^2, z, {\bm q}^2 )  \  R_{g^*q^*}(z, {\bm q}, {\bm k}), 
\end{align}
where $  \hat{\sigma} $  is the following 
off-shell   continuation of the    pointlike  $q\bar{q} \to Z$
matrix element,    
\begin{align}
  \label{eq:sigmahat}
  \hat{\sigma}( x_1 x_2 s, M_Z^2,  z, {\bm q}^2 ) 
&= \sqrt{2} G_F M_Z^2 (V_q^2 + A_q^2)  \frac{\pi}{N_c} \delta(z x_1 x_2 s- {\bm q}^2 - M_Z^2),
\end{align}
with $M_Z$ the mass of the $Z$-boson, $G_F$ the Fermi coupling,  
$V_q$ and $A_q$ the coefficients of the vector and axial couplings of
the $Z$ to a quark, and  
   $ R_{g^*q^*}(z, {\bm q}, {\bm k}) $ is given by 
\begin{align}
  \label{eq:impr_splitt}
 &  R_{g^*q^*}(z, {\bm q}, {\bm k}) = 
\frac{T_R  \alpha_s  }{ 2\pi}
 \frac{z(1-z)}{({\bm \Delta}^2 + z(1-z){\bm k}^2 )^2}  \notag \\
&  \qquad 
\times  \left[  \frac{{\bm \Delta}^2}{z(1-z)} 
+
 4 (1-2 z)  {\bm \Delta}\cdot {\bm k}
 -
 4\frac{({\bm \Delta}\cdot {\bm k})^2  }{{\bm k}^2} + 4 z(1-z) {\bm k}^2  
 \right] , 
\end{align}
where  $T_R = 1 / 2$,  
${\bm \Delta} = {\bm q} - z {\bm k}$.   Upon azimuthal average, 
eq.~\eqref{eq:impr_splitt}
returns  the TMD  gluon-to-quark  splitting 
kernel~\cite{Catani:1994sq}. 
We  examine   azimuthal    effects    numerically  in the next section. 
By replacing $ R_{g^*q^*}(z, {\bm q}, {\bm k}) $ with   its azimuthal average and 
neglecting the  order-$ z {\bm k} $  contribution to the 
 transverse momentum  recoil  in eq.~\eqref{eq:sigmahat}, we get  
\begin{align}
  \label{eq:heXsecbetter}
   \sigma_{qg^* \to Zq}^{k_T-\text{fact.}} &\simeq
 \int\limits_{0}^1 d z \int 
 \frac{d^2 {\bm \Delta} }{\pi  {\bm \Delta}^2 } \; 
   \hat{\sigma} ( x_1 x_2 s, M_Z^2, z, {\bm \Delta}^2 )  \ 
   \Theta\left(   \mu^2_F -    z   {\bm k}^2     -  {{   {\bm \Delta}^2   } \over { 1-z}}  \right) \  
\frac{\alpha_s}{2 \pi} \  {P}_{qg} \left(z, {\bm k}^2 ,  {\bm \Delta}^2 \right)  , 
\end{align}
where 
\begin{align}
  \label{eq:ktsplitt_def}
  {P}_{qg} \left(z, {{\bm k}^2}, {{\bm \Delta}^2} \right) = 
T_R \left( 
            \frac{{\bm \Delta}^2}{{\bm \Delta}^2 + z(1-z){\bm k}^2}
\right)^2 
\left[
{(1-z)^2 + z^2 }
 + 4z^2 (1-z)^2 \frac{{\bm k}^2}{{\bm \Delta}^2}
\right] . 
\end{align}
It was  pointed out 
in~\cite{Catani:1994sq}  that  eq.~\eqref{eq:ktsplitt_def} allows one to     
relate    the sea quark distribution,   with next-to-leading logarithmic 
accuracy at    small $x$,   to   the  small-$x$ gluon Green's function 
$ \mathcal{G}\left( x , {\bm k}^2, {\mu}^2   \right)$ obeying the   BFKL equation. 
 Although evaluated off-shell,  
 the splitting function in  eq.~\eqref{eq:ktsplitt_def}   
 is universal~\cite{Catani:1994sq,cc05}.   
 It reduces to the    
 collinear   splitting function  
at lowest order  for $  {\bm k} = 0$,  
but in the high energy limit  it     factorizes correctly   
  the finite $  {\bm k} $ dependence  to all orders. 
By using this result,  
the unintegrated sea quark distribution  is given by 
\begin{align}
  \label{eq:seaqark}
  \mathcal{Q}^{\text{sea}} \left(x, \frac{{\bm \Delta}^2}{\mu^2}, \frac{\mu_F^2} {\mu^2} \right) &=   
   \int\limits_x^1  \frac{d z}{z}
\int 
 d {\bm k}^2  \  \Theta\left(  \mu_F^2 - \frac{{\bm \Delta}^2 + z (1-z) {\bm k}^2}{1-z}    \right)
\notag \\
& \qquad \qquad \qquad  \qquad  \frac{1}{{\bm \Delta}^2}
  \frac{\alpha_s}{2 \pi} {P}_{qg} \left(z, {\bm k}^2 ,  {\bm \Delta}^2\right) 
\mathcal{G}\left(\frac{x}{z}, {\bm k}^2, {\mu}^2   \right). 
\end{align}
We thus   write  the 
 forward $Z$-boson  hadronic cross section in the high energy factorized form 
\begin{align}
  \label{eq:total_hadronic_kt}
   \sigma_{pp \to Z} (s, M_Z^2)& = 
\sum_j \int \limits_0^1 d x_1  \int  {{  d^2 {\bm \Delta} } \over \pi }  \int \limits_0^1 d x_2 \,
 \hat{\sigma} \ 
\mathcal{Q}^{\text{sea}}   
\left(x_1, \frac{{\bm \Delta}^2}{\mu^2}, \frac{\mu_F^2}{ \mu^2}\right) \  Q^{(j)}\left(x_2, \frac{\mu_F^2}{\mu^2}\right),
\end{align} 
with $\mathcal{Q}^{\text{sea}} $ given in eq.~\eqref{eq:seaqark} and
$Q^{(j)}$ the (integrated) parton distribution as provided by
collinear factorization.  The scale $\mu^2$ in eqs.~\eqref{eq:seaqark}
and~\eqref{eq:total_hadronic_kt} is an initial infra-red scale at
which the parton distribution functions are defined.  The dependence
on this scale occurs for the TMD distributions as a direct consequence
of the matching procedure between high energy and collinear  
factorization. The presence of collinear singularities in the
integrations over momentum lines connecting the 2PI amplitudes 
 requires the use of a regulator for the
BFKL gluon Green's function which breaks the scale invariance of the
LO BFKL equation. In addition to the BFKL gluon density, also the
integral over ${\bm \Delta}^2$ in eq.~\eqref{eq:total_hadronic_kt} is
collinear divergent in the limit ${\bm k}^2 \to 0$ and requires
regularization, leading to an additional dependence on the scale
$\mu^2$.  While in~\cite{Catani:1994sq} this is achieved through
dimensional regularization in $d = 4 + 2 \epsilon$ dimensions, a
cut-off regularization will be used in future Monte-Carlo realizations
of eq.~\eqref{eq:seaqark}. The dependence on the scale $\mu$ 
 cancels up to second order corrections, {\it i.e.} to order  
$\alpha_s (\alpha_s \ln 1/x)^n$ which is the accuracy at which
eqs.~\eqref{eq:seaqark} and \eqref{eq:total_hadronic_kt} are valid.

We will be interested in applications of the sea quark distribution in
eq.~\eqref{eq:seaqark} to shower Monte Carlo generators including the
transverse momentum dependence of the parton branching, such
as~\cite{casc}. Compared to previous TMD
approaches~\cite{kwie03,Watt:2003vf,jadach09}, this distribution
includes the finite-$ {\bm k} $ dependence of the
kernel~\eqref{eq:ktsplitt_def}, responsible for all-order
logarithmically enhanced corrections to quark evolution at small $x$.
For these applications it is relevant to investigate the size of
corrections associated with the high-energy kinematic approximations
discussed earlier.  We turn to this in the next section, and study the
numerical effect of the kinematic contributions on the off-shell
$Z$-boson production cross section.

\section{Kinematic effects in the partonic $ Z$ cross section} 

In this section we discuss that 
  the accuracy of the factorized expression obtained above 
  can be increased  by  going beyond the strong ordering 
  approximation in eqs.~\eqref{eq:Zvertex},\eqref{eq:GammaZ}.    
  We will    write  the convolution integral in terms of the virtuality of      
the four-momentum exchanged in the $t$-channel, $|t| = - \ell^2$, rather   
 than the transverse momentum.      
While the two  formulations coincide in the asymptotic high-energy limit 
  $z \to 0$,      they  differ by  inclusion of   finite-$z$  terms. 
 We     examine  numerically the role  of  these terms.  

The formulation in $|t|$  arises  naturally  if we remove  the
constraint  $\ell \cdot p_1 = 0$  in  the  effective vertex  of  eq.~\eqref{eq:GammaZ}.  The
kinematically improved version of this vertex is
\begin{align}
  \label{eq:Ztilde}
  \tilde{\Gamma}^\mu_+(p, \ell, q) &  = 
   \left( \gamma^\mu + \fdag{\ell} \cdot \frac{p_1^\mu \,}{p_1 \cdot q}   \right) . 
\end{align}
Gauge invariance can be  verified similarly to 
eqs.~\eqref{eq:minusvertex_gen},\eqref{eq:gamma_tilde_regg}. 
One then  recovers 
 eq.~\eqref{eq:heXsec}  with $\hat{\sigma} $   
replaced by
\begin{align}
\label{eq:part_tfac}
  \hat{\sigma}^{t} 
&=
\sqrt{2} G_F M_Z^2 (V_q^2 + A_q^2)  \frac{\pi}{N_c} \delta(z x_1 x_2 s  + t - M_Z^2),
\end{align}
 where 
\begin{align}
  \label{eq:t}
  -t & = - \ell^2 =  \frac{{\bm \Delta}^2}{1-z} + z {\bm k}^2  . 
\end{align}
 The angular average can  now 
be performed exactly, and  yields 
\begin{align}
  \label{eq:heXsecbetterprime0}
  \sigma_{qg^* \to Zq}^{t} &= 
 \int\limits_{0}^1 d z \int \limits_0^\infty \frac{d {\bm \Delta}^2 }{ {\bm \Delta}^2 } \,  {\hat \sigma}^t \  
 \Theta\left(\mu^2_F - \frac{{\bm \Delta}^2 + z(1-z) {\bm k}^2}{1-z} \right)   \  \frac{\alpha_s}{2 \pi}  {P}_{qg} \left(z,  {\bm k}^2 ,  {\bm \Delta}^2 \right).
  \end{align}
  Unlike eq.~\eqref{eq:heXsecbetter} the above expression no longer
  decouples in transverse momentum space.  However, by changing variable  to   $| t | $
  in  the convolution  integral, we get  
\begin{align}
  \label{eq:heXsecbetterprime}
   \sigma_{qg^* \to Zq}^{t}(x_1 x_2 s, {\bm k}^2, M_Z^2) &= 
 \int\limits_{0}^1 d z \int  \frac{d |t| }{ |t|  - z {\bm k}^2}  \,
  \Theta\left(  \mu^2_F - |t|   \right) \  \hat{\sigma}( x_1 x_2 s, M_Z^2,  z,  |t| ) \
 \notag \\
& \qquad  \qquad     \times  \    \Theta\left( |t|  - z   {\bm k}^2   \right)   \ 
 \frac{\alpha_s}{2 \pi}  {P}_{qg} \left( z,  {\bm k}^2 , 
 (1-z)( |t| -z { \bm k}^2)  \right) . 
\end{align}
This  result shares some features with the 
results  in \cite{Watt:2003vf} and
 \cite{Martin:2009ii}. In
particular,  as in~\cite{Martin:2009ii}
it takes into account   subleading corrections to  
 strong ordering   by keeping track of the exact kinematic 
 relation between virtuality and transverse momentum. 
 It was shown in the  numerical NLO-DGLAP analysis of~\cite{Martin:2009ii} 
 that  these kinematic   contributions  provide  a large fraction of the full  corrections at NLO. 
On the other hand, unlike~\cite{Watt:2003vf,Martin:2009ii}  
eq.~\eqref{eq:heXsecbetterprime}    contains  the  dynamical  effects of 
transverse momentum dependent terms in the gluon-to-quark splitting kernel.   
These become potentially important in the forward production kinematics, 
as they give rise to  logarithmically enhanced terms for  small $x$ at higher orders in 
$\alpha_s$.  

While the general structure
of the unintegrated sea quark density eq.~\eqref{eq:seaqark} remains
unchanged, the  off-shellness is now expressed in terms of the  absolute
value of the four momentum square exchanged in the $t$-channel,
\begin{align}
  \label{eq:seaqark-t}
  \mathcal{Q}^{\text{sea}}_{t} \left(x, \frac{|t|}{\mu^2}, \frac{\mu^2_F}{\mu^2}\right) &=   
   \int\limits_x^1  \frac{d z}{z} 
\int 
 d {\bm k}^2   
 \    \Theta\left( |t|  - z   {\bm k}^2   \right)   
  \        {P}_{qg} \left( z, {\bm k}^2 ,  (1-z)( |t| -z { \bm k}^2 ) \right)  
\notag \\
& \qquad \qquad \qquad \qquad   
 \times \ 
  \frac{1}{|t| - z {\bm k}^2}
  \frac{\alpha_s}{2 \pi} 
\mathcal{G}\left(\frac{x}{z}, {\bm k}^2, {\mu}^2   \right). 
\end{align}
   In~\cite{Catani:1994sq,cc05}   the low-$  {\bm k}^2 $  
 behavior of eqs.~\eqref{eq:seaqark},\eqref{eq:seaqark-t} is 
 analyzed using dimensional regularization.  For   the parton-shower 
  applications~\cite{prepar}   an infrared cut-off is applied.   In the numerical  
  study that follows, we  examine the effect of the TMD sea quark distribution 
  on the  $Z$-boson   partonic cross section.  For this purpose  we will work 
 at fixed  $  {\bm k}^2 $. 

\begin{figure}[h!]
  \centering
  \parbox{.8 \textwidth }{\includegraphics[width = .7 \textwidth ]{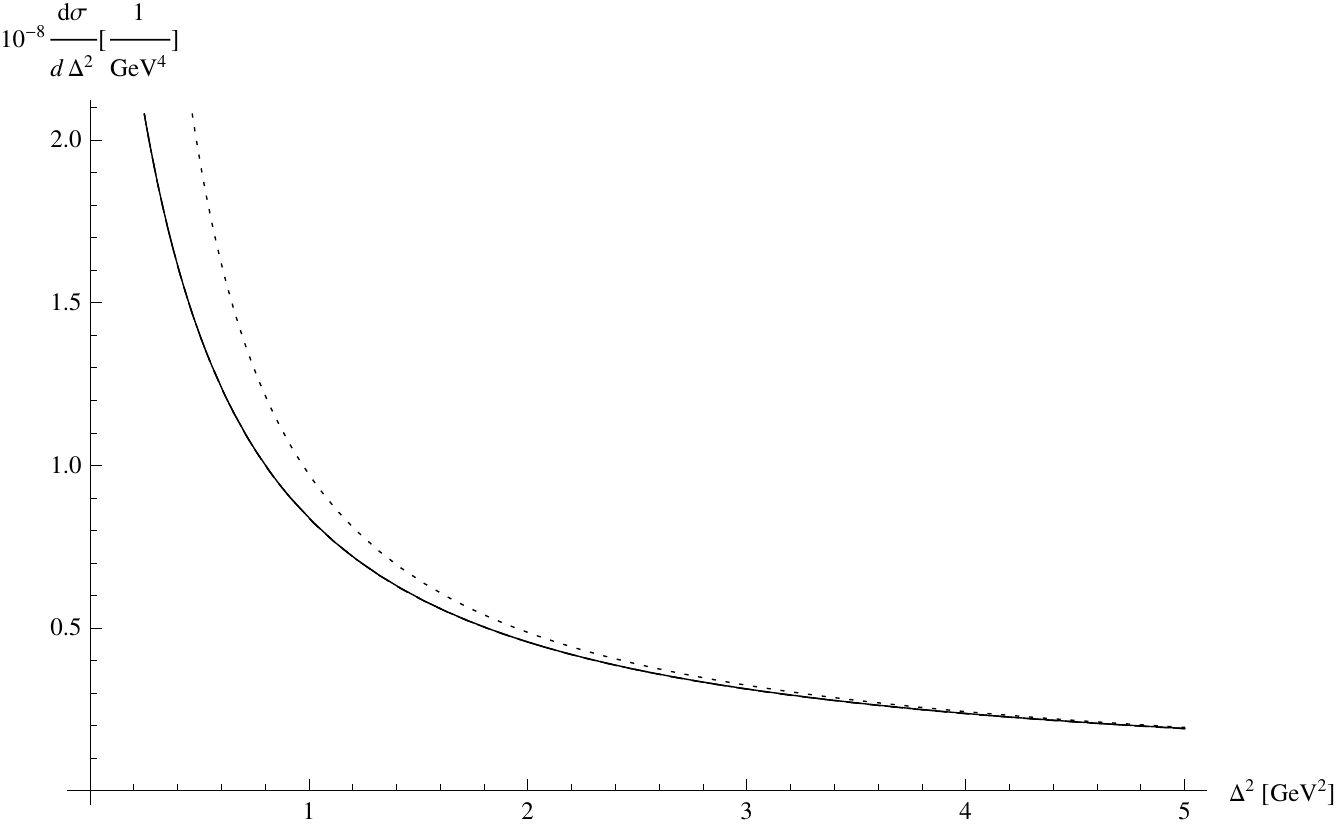}}   \parbox{.8 \textwidth}{\center (a)} \\
\parbox{.8 \textwidth }{\includegraphics[width = .7 \textwidth ]{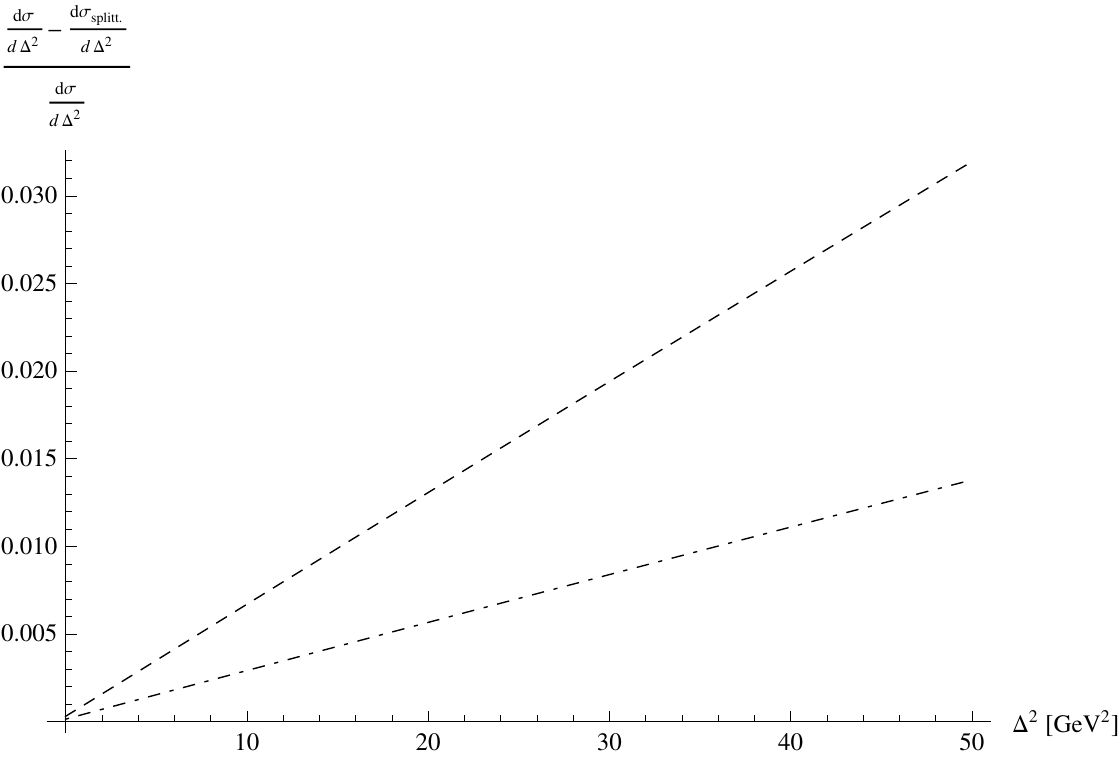}}   \parbox{.8 \textwidth}{\center (b)} \\
 \caption{(a):    ${\bm \Delta}^2 $  dependence of    the   
   differential cross section  $ d\sigma /  d {\bm \Delta}^2 $ for   small $| {\bm \Delta} | $: (solid) full; (dashed) no plus-momentum ordering; (dot-dashed) no plus-momentum and minus-momentum ordering;  (dotted)  collinear approximation.  All but the last curve overlap in this region.  We set 
 $x_1 x_2 s = 2.5 M_Z^2 $, ${\bm k}^2 = 2$ GeV$^2$.  (b): Relative deviations  in   
 the differential cross section $ d\sigma /  d {\bm \Delta}^2 $:  (dashed) no plus-momentum ordering; 
 (dot-dashed)    no plus-momentum and minus-momentum ordering.}
 \label{fig:deltasmall}
\end{figure}

\begin{figure}[h!]
  \centering
\parbox{.8\textwidth }{\includegraphics[width = .8 \textwidth ]{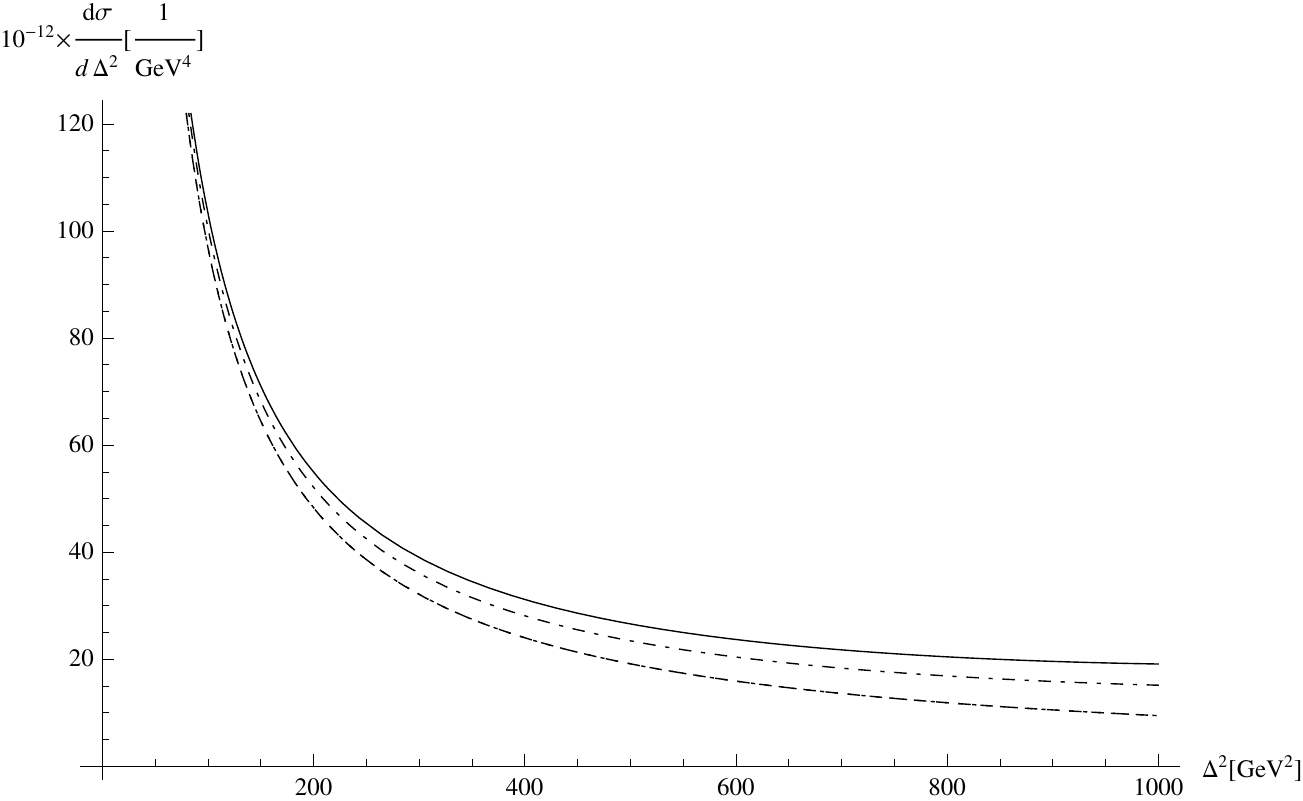}} \\
\parbox{.8\textwidth }{\center (a)}
\parbox{.8\textwidth }{\includegraphics[width = .8 \textwidth ]{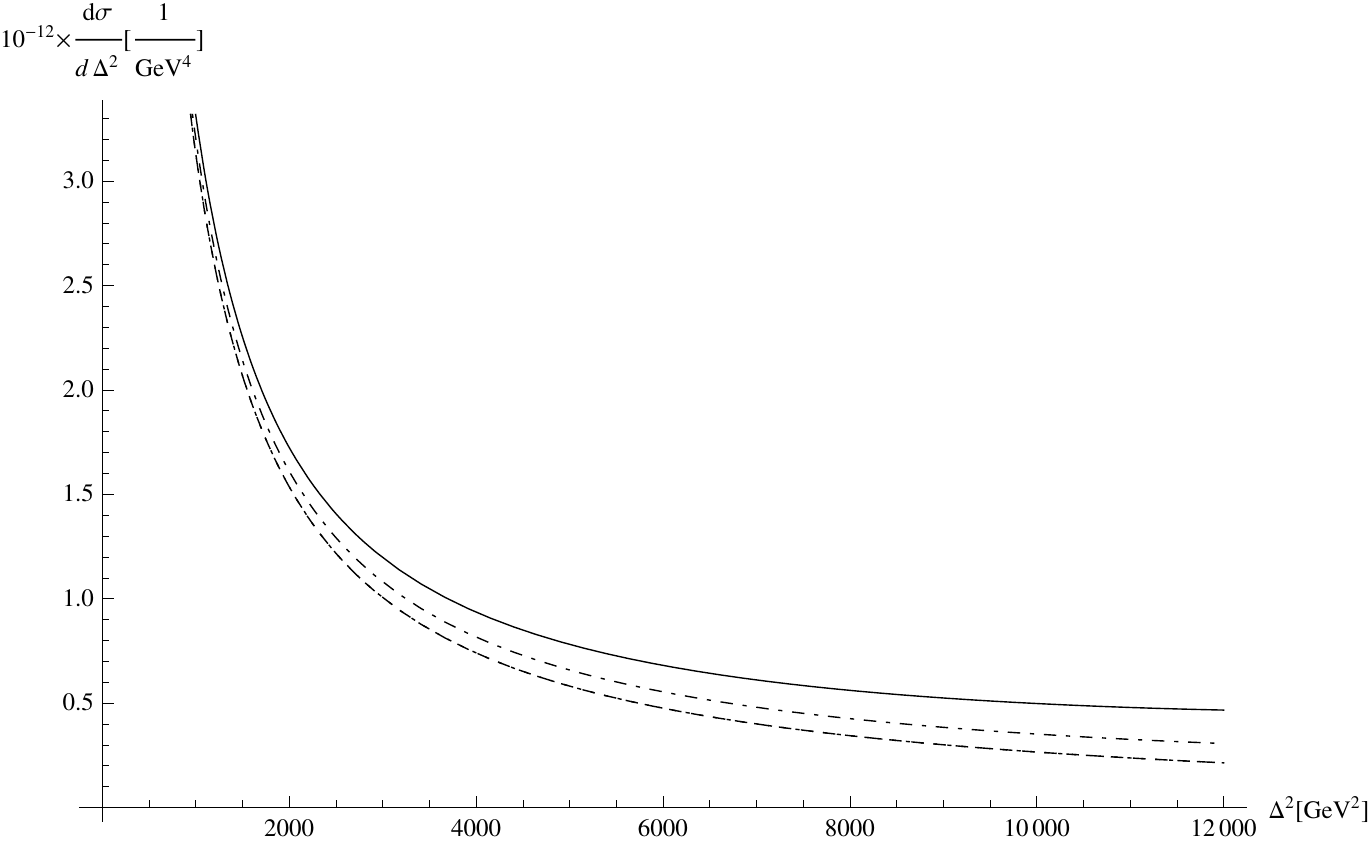}} \\
\parbox{.8\textwidth }{\center (b)}
   \caption{$ {\bm \Delta}^2 $  dependence of 
   the   differential cross section $ d\sigma /  d {\bm \Delta}^2 $ 
 in  the  large $| {\bm \Delta} | $ region:   (solid)  full; 
 (dashed) no plus-momentum ordering;  
 (dot-dashed)    no plus-momentum and minus-momentum ordering; 
  (dotted)  collinear approximation. 
  We set ${\bm k}^2 = 2$ GeV$^2$; 
(a)  $x_1 x_2 s = 2.5 M_Z^2 $,   (b) $x_1 x_2 s = 12.5 M_Z^2 $.  }
 \label{fig:deltabigg}
\end{figure}

 We    perform a  numerical   comparison of 
 eq.~\eqref{eq:heXsecbetter} and eq.~\eqref{eq:heXsecbetterprime}, 
 corresponding respectively   to the  kinematic effects of   going beyond 
   the plus-momentum  strong ordering 
and  the minus-momentum  strong ordering   
in the high-energy factorized formula. 
 We compare this    with the 
  full $qg^* \to qZ$ matrix element result.
We consider also the result  of the collinear approximation obtained 
 by setting ${\bm k}^2 = 0$ in eq.~\eqref{eq:heXsecbetter}.    
  We examine the dependence on the transverse momentum variable 
$ {\bm \Delta}  =  {\bm q} - z {\bm k}  $.  Here $ {\bm q} $  is the vector boson 
transverse momentum;  $z$ is the  plus-momentum fraction transferred 
 to the vector boson from the incoming gluon that results   from 
 the small-$x$ initial-state  shower;  $ {\bm k} $ is the  transverse 
 momentum carried  by the partons emitted in this  shower (i.e.,  partons  
 radiated in addition to the leading  quark  against which the vector boson 
 recoils).   In fig.~\ref{fig:deltasmall} and fig.~\ref{fig:deltabigg} we examine  
 respectively the 
 region where $| {\bm \Delta} | $ is  small compared to  the $Z$ boson mass $M_Z$ 
  and the region where $| {\bm \Delta} | $ is on   the order of the  
  $ Z$ boson mass or larger.  


  For  small  $| {\bm \Delta} | $
we find that   the differences between 
eq.~\eqref{eq:heXsecbetter} and eq.~\eqref{eq:heXsecbetterprime} are 
numerically small,  and  that  both expressions are   close to  the full result;   as 
  $| {\bm \Delta} | $ increases, we find that    the deviations   due to the 
  kinematic contributions by which 
eq.~\eqref{eq:heXsecbetter} and eq.~\eqref{eq:heXsecbetterprime}   differ 
    become non-negligible,  and that 
eq.~\eqref{eq:heXsecbetterprime} gives    a   better approximation    to the full result. 
Fig.~\ref{fig:deltasmall}  illustrates the small  $| {\bm \Delta} | $ region.  
 In fig.~\ref{fig:deltasmall}(a) we plot the differential cross section 
 $ d \sigma / d   {\bm \Delta}^2 $ for fixed values of  ${\bm k}^2$ and $x_1 x_2 s$. 
 The solid, dashed and dot-dashed curves overlap  
 on  the scale of this plot, while the dotted curve, corresponding to the 
collinear  approximation, deviates 
from them when $| {\bm \Delta} |  \lesssim   | {\bm k} | $.     
    In fig.~\ref{fig:deltasmall}(b) 
we zoom in on the  small 
relative  deviations 
of eq.~\eqref{eq:heXsecbetter} and eq.~\eqref{eq:heXsecbetterprime} 
from  the full result  for the differential cross section 
 $ d \sigma / d   {\bm \Delta}^2 $.  We see  that   throughout the range shown 
 in fig.~\ref{fig:deltasmall}(b) 
the deviations  are  at most of  the order  of  few    percent.  
In this region the factorized expressions  based on 
 reggeized quark graphs 
 are close to the  full result, 
  independently   of  
 the  different kinematic  approximations on  the longitudinal momentum orderings. 
 What dominates  this region 
 are  the transverse momentum dependent terms, which are kept 
 correctly by 
  both eq.~\eqref{eq:heXsecbetter} and eq.~\eqref{eq:heXsecbetterprime}. 
 On the other hand, 
 we see  
 in  fig.~\ref{fig:deltasmall}(a) 
  that in the  small  $| {\bm \Delta} | $  region   the collinear  expression obtained 
by setting ${\bm k}^2 = 0$ in eq.~\eqref{eq:heXsecbetter} 
is far from the full result, due to missing  TMD  corrections to  the 
kernel~\eqref{eq:ktsplitt_def}.  

As 
  $| {\bm \Delta} | $ increases the kinematic effects  from   
   the plus-momentum   ordering 
and   minus-momentum   ordering   
  become significant.    
Fig.~\ref{fig:deltabigg}  
illustrates 
these effects 
for $| {\bm \Delta} | $    
 on   the order of the    $ Z $ boson mass $M_Z$,  or larger. 
Note that in this region the 
collinear ${\bm k}^2 = 0$ expression gives a good approximation 
to eq.~\eqref{eq:heXsecbetter}.  The dashed and dotted curves nearly overlap 
on the scale of the plots in fig.~\ref{fig:deltabigg}. 
However, the full result contains significant corrections with respect to these curves, 
corresponding to terms  that are subdominant  
both in the collinear and in the high-energy expansions. 
We see that a   non-negligible part 
 of these corrections   can be  taken into account by  including 
kinematic  contributions  from  terms beyond  the strong ordering in the lightcone 
minus momenta, as   
eq.~\eqref{eq:heXsecbetterprime}   does. 
This is analogous to   effects observed in~\cite{Martin:2009ii}. 
As a result    
eq.~\eqref{eq:heXsecbetterprime} 
provides a closer approximation to the  full matrix element. 
It can  be verified numerically that  the corrections die out in the limit of 
asymptotically  large $s$.  This  analysis  is however significant 
for applications to  finite collider energies. 

We  thus   observe that  
taking into account 
 TMD  terms     in 
the parton splitting and the reggeized quark coefficient   
enables one 
to  extend the  
description of vector boson production 
in terms of 
factorized  quark distributions 
 from  the collinear region  into   the region where 
 $| {\bm \Delta} | $   is of   order   $     | {\bm k} | $ and 
 transverse momenta are no longer  strongly  ordered.  
This is a region where collinear approximations are seen not to be  sufficient, and  it 
is relevant to  forward Drell-Yan.  We will use the unintegrated quark distribution thus 
defined in forthcoming studies~\cite{prepar}.    
As a result of  TMD terms,  on the other hand, 
the  initial state kinematics  becomes  rather more complex.  
Corrections to  longitudinal momentum     ordering   in  the effective vertices   
are seen to be  numerically  significant when  $| {\bm \Delta} | $  is of order 
the vector boson mass.  To this end  it is  useful  to 
employ   the formulation given in this section 
which includes  the exact one-loop kinematics.

\section{Conclusion}

Drell-Yan production processes are instrumental in   hadron  collider experiments 
 both for  hadronic physics studies and for new physics searches. 
At the LHC   a   new kinematic region   opens up 
for   forward Drell-Yan, in which     the QCD treatment of the production 
 process and  associated 
final states    acquires new features  due to   multi-parton   radiation  
over long rapidity intervals.    The  emission 
of soft gluons that are not  collinearly  ordered    becomes  
relevant, and effects of color coherence set in 
associated  with  the region of small longitudinal momentum 
fractions $x$. 
Taking this into  account calls  for  improved formulations  of 
 factorization in QCD  at  unintegrated, or transverse momentum dependent,  level.

 Fully  general TMD 
 factorization formulas are still lacking~\cite{jcc-book,mert-rog}, due to the 
 difficulty in disentangling  systematically  soft and collinear gluon 
correlations between initial and final states.   
This is  illustrated,  for instance,   by   
 back-to-back  di-hadron and  di-jet 
 hadroproduction~\cite{muld-rog,nonfac07}.   
In the case of small $x$, however,  TMD factorization results  
exist~\cite{Lipatov:1995pn,ktfac} based on   
dominance of single gluon helicity at high energy.  
These  results  can be  
used to define  gauge-invariantly the TMD gluon 
distribution.    See     recent works  in~\cite{avsar11,mue11,qiuetal11}.      
In this paper we have  used the method~\cite{Catani:1994sq}   
 to extend  the above   treatment 
to the  TMD sea-quark distribution.  This 
is the dominant  channel coupling  to  forward  Drell-Yan production. 
We have related the  TMD splitting  kernel~\cite{Catani:1994sq} to the forward 
$Z$-boson  cross section.  The reggeized quark  
formalism~\cite{Bogdan:2006af,
  Lipatov:2000se} is used to treat the off-shell quark coefficient  
  and ensure  gauge invariance. 
  We have analyzed numerically   effects  
  due to  the transverse momentum  kinematics 
  in the relationship  between the $Z$ boson spectrum and the TMD sea-quark 
  distribution. 

Phenomenological applications of the reggeized quark formalism  were 
investigated in~\cite{knie-sale}  using  
the approach~\cite{kimber,Watt:2003vf} to TMD pdfs.  
Compared to   these  
studies,  the main feature of  our   approach  
  is that it includes, besides  the 
  lowest-order  quark  splitting function,      
    the full  series of    finite-$  k_\perp $  terms in the TMD kernel.  
  These terms  control  the perturbative summation of 
   small-$x$ logarithmic corrections to flavor-singlet 
 observables    to all orders in   $\alpha_s$.

 Future  extensions  of  the   
  results  above  involve  
  several directions.  One concerns  large-$x$  corrections, likely to be important for 
  Drell-Yan phenomenology.    See~\cite{idi-sci,ceccopie,chered,jccfh01}  
 for   discussion   of   
  $x \to 1$  issues  in TMD  quark distributions.  
   Another direction concerns      nonlinear effects 
  from     high  parton densities in the small $x$  region.  Recent 
  works~\cite{doming11,xiao} 
  study   multiple scattering contributions in  
 dense targets and  nuclei.  
 Techniques such as~\cite{s-channel}   have been proposed 
to  incorporate 
 the treatment of     multiple-gluon rescattering graphs  at small $x$   
starting from  the operator matrix 
elements~\cite{jcc-book,mert-rog}  for parton distributions. 
They may    be  helpful    for   extensions  to  the high density region 
that are aimed at   retaining   accuracy  also in the 
treatment of  contributions from high p$_{\rm{T}}$ 
processes, e.g.\  Drell-Yan on nuclei~\cite{salg}.

The results  in this paper can  be implemented in a parton shower Monte Carlo 
generator including  transverse-momentum dependent 
 branching,  such as~\cite{casc}.  
 Work along these lines is in progress~\cite{prepar}. 
This will allow one  to address  predictions for exclusive final-state observables 
associated with forward Drell-Yan production, for which intense  experimental 
 activity is forthcoming~\cite{lhc-b-dy}.

\vskip 1 cm

\noindent {\bf Acknowledgments}. 
M.~H.\  is grateful for financial support from the German Academic
Exchange Service (DAAD), the MICINN under grant FPA2010-17747, the
Research Executive Agency (REA) of the European Union under the Grant
Agreement number PITN-GA-2010-264564 (LHCPhenoNet) and the Helmholtz
Terascale Analysis Center.   F.~H.\     thanks   the CERN 
Theory Division    for    hospitality and support.

\end{document}